\documentclass[submission,copyright,creativecommons]{eptcs}
\providecommand{\event}{Workshop on Formal Methods for Autonomous Systems (FMAS 2021)} % Name of the event you are submitting to
\usepackage{comment}
\usepackage{url}

\usepackage{enumitem}
\usepackage{subfig}
\usepackage{multirow}
\usepackage[table]{xcolor}
\usepackage{tikz}

\newtheorem{theorem}{Theorem}

\usepackage{hyperref}
\usepackage{cleveref}

\crefname{section}{Sec.}{Sec.}
\Crefname{section}{Section}{Sections}
\crefname{subsection}{Sec.}{Sec.}
\Crefname{subsection}{Section}{Sections}
\crefname{subsubsection}{Sec.}{Sec.}
\Crefname{subsubsection}{Section}{Sections}

\crefname{theorem}{Thm.}{Thm.}
\Crefname{theorem}{Theorem}{Theorems}
\crefname{lemma}{Lem.}{Lem.}
\Crefname{lemma}{Lemma}{Lemmata}
\crefname{corollary}{Cor.}{Cor.}
\Crefname{corollary}{Corollary}{Corollaries}
\crefname{proposition}{Prop.}{Prop.}
\Crefname{proposition}{Proposition}{Propositions}
\crefname{fact}{Fact}{Facts}
\Crefname{fact}{Fact}{Facts}
\crefname{definition}{Def.}{Def.}
\Crefname{definition}{Definition}{Definitions}

\crefname{example}{Ex.}{Ex.}
\Crefname{example}{Example}{Examples}
\crefname{remark}{Rm.}{Rm.}
\Crefname{remark}{Remark}{Remarks}
\crefname{equation}{Eq.}{Eq.}
\Crefname{equation}{Equation}{Equations}

\crefname{figure}{Fig.}{Fig.}
\Crefname{figure}{Figure}{Figures}
\crefname{table}{Tab.}{Tab.}
\Crefname{table}{Table}{Tables}
\crefname{listing}{\lstlistingname}{\lstlistingname}
\Crefname{listing}{Listing}{Listings}

\title{Online Strategy Synthesis for Safe and Optimized Control of Steerable Needles}

\author{
Sascha Lehmann$^{1}$ \qquad Antje Rogalla$^{1}$ \qquad Maximilian Neidhardt$^{2}$ \\
\qquad Alexander Schlaefer$^{2}$ \qquad Sibylle Schupp$^{1}$
\institute{
$^{1}$Institute for Software Systems
\qquad
$^{2}$Institute of Medical Technology\\Hamburg University of Technology, Hamburg, Germany\thanks{This study was partially funded by the TUHH i$^3$~lab initiative (T-LP-E01-WTM-1801-02), DFG SCHU 2479, and DFG SCHL 1844/6-1.}}
\email{\{s.lehmann, antje.rogalla, maximilian.neidhardt, schlaefer, schupp\}@tuhh.de}
}

\def\titlerunning{Safe and Optimized Strategy Synthesis}
\def\authorrunning{S. Lehmann et al.}

\begin{document}
\maketitle

%%%%%%%%%%%%%%%%%%%%%%%%%%%%%%%%%%%%%%%%%%%%%%%%%%%%%%%%%%%%%%%%%%%%%%%%%%%%%%%%
%%% Raw Abstract %%%
%%%%%%%%%%%%%%%%%%%%%%%%%%%%%%%%%%%%%%%%%%%%%%%%%%%%%%%%%%%%%%%%%%%%%%%%%%%%%%%%

% Autonomous systems are often applied in uncertain environments, which require prospective action planning and retrospective data evaluation for future planning to ensure safe operation. Formal approaches may support these systems with safety guarantees, but are usually expensive and do not scale well with growing system complexity. In this paper, we introduce online strategy synthesis based on classical strategy synthesis to derive formal safety guarantees while reacting and adapting to environment changes. To guarantee safety online, we split the environment into region types which determine the acceptance of action plans and trigger local correcting actions. Using model checking on a frequently updated model, we can then derive locally safe action plans (prospectively), and match the current model against new observations via reachability checks (retrospectively). As use case, we successfully apply online strategy synthesis to medical needle steering, i.e., navigating a (flexible and beveled) needle through tissue towards a target without damaging its surroundings.

%%%%%%%%%%%%%%%%%%%%%%%%%%%%%%%%%%%%%%%%%%%%%%%%%%%%%%%%%%%%%%%%%%%%%%%%%%%%%%%%
%%% Abstract %%%
%%%%%%%%%%%%%%%%%%%%%%%%%%%%%%%%%%%%%%%%%%%%%%%%%%%%%%%%%%%%%%%%%%%%%%%%%%%%%%%%
\begin{abstract}
  Autonomous systems are often applied in uncertain environments, which require prospective action planning and retrospective data evaluation for future planning to ensure safe operation.
  Formal approaches may support these systems with safety guarantees, but are usually expensive and do not scale well with growing system complexity.
  In this paper, we introduce \textit{online strategy synthesis} based on classical strategy synthesis to derive formal safety guarantees while reacting and adapting to environment changes.
  To guarantee safety online, we split the environment into region types which determine the acceptance of action plans and trigger local correcting actions.
  Using model checking on a frequently updated model, we can then derive locally safe action plans (prospectively), and match the current model against new observations via reachability checks (retrospectively).
  As use case, we successfully apply online strategy synthesis to medical needle steering, i.e., navigating a (flexible and beveled) needle through tissue towards a target without damaging its surroundings.
\end{abstract}

%%%%%%%%%%%%%%%%%%%%%%%%%%%%%%%%%%%%%%%%%%%%%%%%%%%%%%%%%%%%%%%%%%%%%%%%%%%%%%%%
%%% Introduction %%%
%%%%%%%%%%%%%%%%%%%%%%%%%%%%%%%%%%%%%%%%%%%%%%%%%%%%%%%%%%%%%%%%%%%%%%%%%%%%%%%%
\section{Introduction} \label{sec:introduction}

In cyber-physical systems (CPS), we deal with the common task of directing a controllable entity through its environment towards a target state defined by local or global goals, without violating frame properties imposed by safety requirements.
For applications of limited complexity with directly measurable environmental parameters (e.g., thermostats), practical control solutions already exist \cite{LuiY2017}.
However, the environment of complex systems is often uncertain, changing, and not entirely measurable, rendering a comprehensive model of such environments for safety prediction infeasible.
Deriving safety guarantees via formal methods imposes further limitations on feasible model complexities.
Using a partial model instead, global safety guarantees might not hold anymore as soon as reality and model deviate.

In practice, safety is only required for the actual trace of the real system.
An \textit{online} approach based on model updates on the fly splits the task of deriving global guarantees into a series of safety verifications on locally valid models with limited scope.
That way, we guarantee safety for all possible near futures of the current system state.
However, three main problems persist:
First, one still needs to guarantee that the system does not reach critical sections due to local deviations of the predicting model.
Second, one needs to decide when to adapt the model or the real system.
Third, among the locally safe solutions, one should choose those that keep the potential number of adaptations low.
We approach these problems by splitting the environment into discrete regions, i.e., unknown, target, safe, critical, and detection regions, based on which we can derive safety-preserving entity actions and perform cost optimizations.

Overall, we approach the safety problem of dynamic CPS by the following four contributions:
\begin{enumerate}[noitemsep, leftmargin=*]
  \item We introduce online strategy synthesis (OnSS) based on frequent model updates and local synthesis
  \item We introduce a region interpretation of the environment (for the real system and model) to guarantee safety in the OnSS workflow
  \item We investigate options of optimal action plan choice from synthesized strategies
  \item We apply OnSS and the region interpretation to the application of autonomous medical needle steering
\end{enumerate}

The remaining paper is structured as follows:
We provide preliminary definitions and related work in \cref{sec:preliminaries}.
Then, in \cref{sec:online-strategy-synthesis}, we discuss the modeling and workflow, safety and optimization aspects of OnSS.
Afterwards, we perform the needle steering experiments in \cref{sec:experiments}, and conclude our work in \cref{sec:conclusion}.

%%%%%%%%%%%%%%%%%%%%%%%%%%%%%%%%%%%%%%%%%%%%%%%%%%%%%%%%%%%%%%%%%%%%%%%%%%%%%%%%
%%% Preliminaries %%%
%%%%%%%%%%%%%%%%%%%%%%%%%%%%%%%%%%%%%%%%%%%%%%%%%%%%%%%%%%%%%%%%%%%%%%%%%%%%%%%%
\section{Preliminaries and Related Work} \label{sec:preliminaries}

A \textbf{game} is a mathematical model of interaction between different decision makers.
A common setting is that of two (usually one \textit{controllable} and one \textit{uncontrollable}) adversarial players with alternating turns.
One established use case is a game of a controllable user against an uncontrollable environment;
in needle steering, these entities are the needle and the tissue, respectively.
Evaluating a game allows determining solutions for winning or losing of a particular party involved.

A \textbf{timed automaton (TA)} is a finite-state machine extended by (real-valued) clocks \cite{Alur1994}.
It is a simple type of hybrid system, with only one differential equation, i.e., a constant change rate for all clocks.
While clocks usually provide a notion of time (where clock invariants in locations and clock guards and reset on edges determine the timed transitions between states), they can be used to model continuous variables in general, including physical variables and accumulated costs.

A \textbf{timed game (TG)} extends the classical timed automaton with a notion of controllable and uncontrollable transitions.
That way, one can model adversarial games between a controller and a (stochastic) environment in timed systems \cite{Bertrand2012}\cite{Bouyer2009}.
The combined formalism applies well to autonomous systems in particular, as they perform actions over time in a (partially) unknown and reactive environment.

\textbf{Offline strategy synthesis (OffSS)} is the automated approach of deriving a \textit{winning strategy}, i.e., a strategy satisfying a given target property \cite{Asarin1995}\cite{Cassez2005}\cite{Maler1995}.
Strategy synthesis on timed games in particular is supported, e.g., by \textit{Uppaal Stratego} \cite{David2015}.
There, a strategy consists of concrete decisions on actions in each state with controllable outgoing transitions.
In needle steering, these decision points are motion actions, i.e., the push, rotate, and pull actions.
OffSS provides the base for our online synthesis approach.

%%%%%%%%%%%%%%%%%%%%%%%%%%%%%%%%%%%%%%%%%%%%%%%%%%%%%%%%%%%%%%%%%%%%%%%%%%%%%%%%
%%% Related Work %%%
%%%%%%%%%%%%%%%%%%%%%%%%%%%%%%%%%%%%%%%%%%%%%%%%%%%%%%%%%%%%%%%%%%%%%%%%%%%%%%%%
Our problem is an instance of controller synthesis and verification of timed games with partial information (about the environment).
Environmental abstractions in static models inevitably lead to incomplete representations of real systems and possibly wrong system behavior at run time \cite{Ferrando2021}.
Several approaches therefore exist that consider environmental uncertainties already during the modeling phase.
\cite{Cassez2007}\cite{David2009} present a transformation of a timed game with partial information into a timed game with complete information,
\cite{Finkbeiner2012} introduces a template-based controller synthesis approach based on automatic abstraction refinement,
and \cite{Bacci2021} presents a new framework for synthesizing strategies for weighted timed games \cite{Bouyer2004} with uncertainties restricted to weights.
Other work introduces algebraic frameworks (e.g., based on risk factors \cite{Gleirscher2021}) for the design of correct safety controllers.
Tools for static controller synthesis and verification include Kronos \cite{Daws1996}, FlySynth \cite{Altisen2002}, Synthia \cite{Peter2011}, and the tool suite UPPAAL \cite{Larsen2018}.
Timed controller synthesis and verification has been applied, among others, to online floor heating \cite{Larsen2016} and vehicle rerouting \cite{Bischopink2020}\cite{Bilgram2021}.
The latter involves uncertain traffic volume and periodically regenerates new strategies based on current traffic data.
Our approach is similar but applies to cognitive or physiological uncertainties.
Specifically, we transfer the static model of timed game needle steering \cite{Rogalla2020} based on a nonholonomic model \cite{Webster2006} into a model that is adapted online.

%%%%%%%%%%%%%%%%%%%%%%%%%%%%%%%%%%%%%%%%%%%%%%%%%%%%%%%%%%%%%%%%%%%%%%%%%%%%%%%%
%%% Optimized Strategy Synthesis %%%
%%%%%%%%%%%%%%%%%%%%%%%%%%%%%%%%%%%%%%%%%%%%%%%%%%%%%%%%%%%%%%%%%%%%%%%%%%%%%%%%
\section{Online Strategy Synthesis}\label{sec:online-strategy-synthesis}
\label{subsec:model-and-workflow}
For online strategy synthesis, we use a single model combining the processes of synthesizing strategies and matching observation data.
The general model consists of five components:

\vspace{6pt}
\noindent
\begin{tabular}{@{}ll}
  1. \textbf{Decision Maker}    & The decision maker who gives instructions to the controlled device \\
  2. \textbf{Controlled Device} & The entity controlled by the decision maker \\
  3. \textbf{Environment}       & The uncontrollable environment \\
  4. \textbf{State Checker}     & The acceptor model which checks properties on individual states \\
  5. \textbf{Data Matcher}      & The acceptor model which matches observations against the action plan
\end{tabular}
\vspace{6pt}

\noindent
The \texttt{Decision Maker}, \texttt{Controlled Device}, and \texttt{Environment} implement the concrete system entities, the \texttt{State Checker} accepts or rejects paths during strategy synthesis, and the \texttt{Data Matcher} validates the correctness of the current prediction model with observation data.
Note that for autonomous devices, the \texttt{Decision Maker} and \texttt{Controlled Device} can be merged into a single \texttt{Actor}.

In needle steering, the \texttt{Decision Maker} is the surgeon (manual) or a stochastic action selector (autonomous), the \texttt{Controlled Device} is the needle, and the \texttt{Environment} is the tissue.
The \texttt{State Checker} checks if critical states are entered, and whether the target is still reachable or already reached, and the \texttt{Data Matcher} matches the observed needle position data against the model-predicted motion.

\begin{figure}[t]
\includegraphics[width=\textwidth]{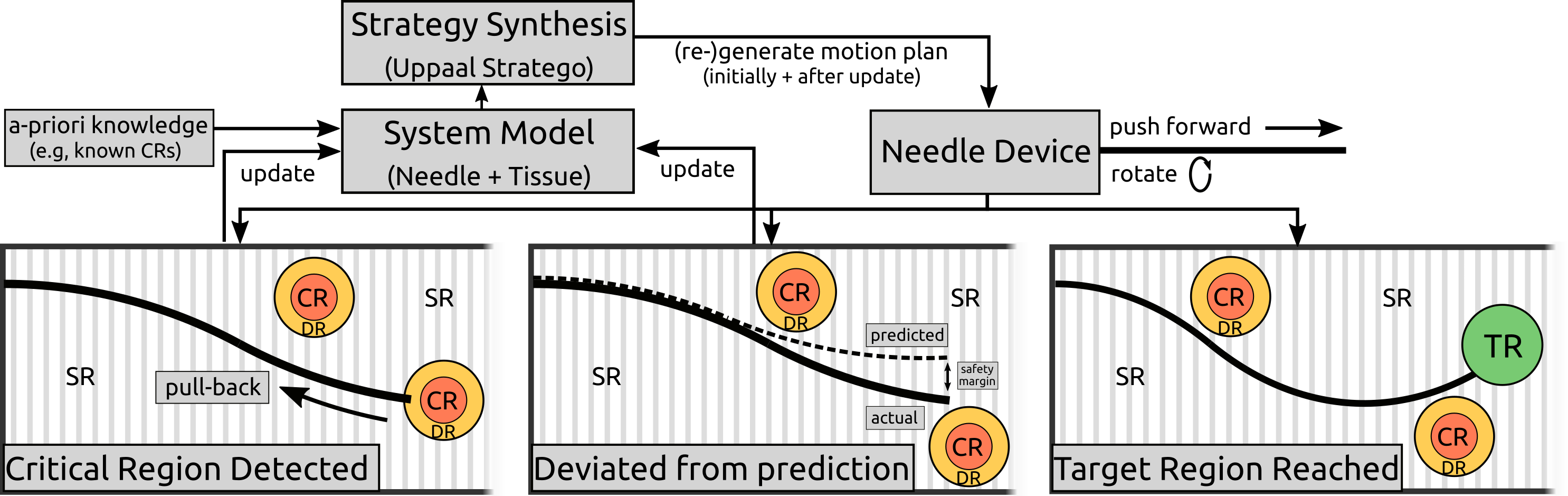}
\caption{The workflow of the online needle steering application.}
\label{fig:needle-steering-workflow}
\end{figure}

In a game setting, the environment is especially important as it determines the safety and reachability of individual system states.
It further determines when and to which extent particular characteristics can be measured.
In an online setting, we need to know in which sections re-evaluation or adaptation of the model or real system is required to still ensure safe operation.
Therefore, we split the complete state space into five regions based on the environment characteristics (see \cref{fig:needle-steering-workflow}).

\vspace{6pt}
\noindent
\begin{tabular}{@{}ll}
  1. \textbf{Unknown regions (UR)}    & Regions not yet classified via a-priori knowledge or discovery \\
  2. \textbf{Safe regions (SR)}       & Regions which do not violate safety \\
  3. \textbf{Critical regions (CR)}   & Regions which violate safety \\
  4. \textbf{Detection regions (DR)}  & Transitional regions between SRs and CRs where upcoming CRs \\
                                      & (= safety violations) can be detected \\
  5. \textbf{Target regions (TR)}     & Regions which we want to reach with the controllable entity
\end{tabular}
\vspace{6pt}

\noindent
In needle steering, the regions map to spatial areas, i.e., the SRs are the uncritical tissue areas, the CRs are hardened tissue and organs, the DRs are pre-rupture deformation sections at which force increases steadily, and the TR is the targeted placement position.

Based on the system model and the regions, the workflow of online strategy synthesis is shown in \cref{fig:needle-steering-workflow}.
Generally speaking, the task is to reach a TR via SRs (or URs if knowledge on SRs is missing) without entering any CRs, which are early detected in surrounding DRs.
The concrete workflow is as follows:
First, the model is initialized (\texttt{Controlled Device}, \texttt{Environment}) with data of the real system's starting state.
Then, an initial strategy is synthesized via reachability check on \texttt{State Checker}.
Up to this point, the approach works offline.
Afterwards, the system is executed and its tracked trace matched against the current motion plan model via \texttt{Data Matcher}.
If an exceeding deviation is detected, the model is again updated, and a new strategy is synthesized.
If no strategy can be found, or if a DR is reached, the system is \textit{readjusted}, i.e., rolled back to a previously visited and known state.
In needle steering, a local pullback of the needle is used for readjustment.
Furthermore, the model is updated to the new system state after readjustment.
If the initial state is re-reached via rollbacks, and again no strategy is found, the process is aborted (and may be started anew under different conditions).
If the target region is reached, the process succeeded.

\vspace{-6pt}
\paragraph{Online Safety Guarantees}\label{subsec:safety}
The online strategy synthesis combines safety and reachability requirements
 but prioritizes safety of the system over reaching a TR.
We assume that every CR has been constructed so that it is surrounded by a DR. For each CR, we define a safe margin, e.g., a minimum required distance so that the CR can be detected early enough and that proper reaction in its DR is possible.
The region sizes are computed based on the underlying system; for safety guarantees, it is necessary that the DRs are larger than the measurable step size under the given time resolution to ensure to halt the system in time when detection-related parameters change.

\begin{figure}[b]
  \newcommand\gr{\cellcolor{green}}
  \newcommand\bk{\cellcolor{black}}
  \newcommand\gy{\cellcolor{black!20}}
  \newcommand\bl{\cellcolor{blue}}
  \newcommand\rd{\cellcolor{red}}

  \small
  \centering
  \begin{minipage}{0.42\textwidth}
    \begin{tabular}[t]{cccccc}
      \hline
      &    & \multicolumn{4}{c}{Real System} \\
      &    & SR & CR & DR & TR \\
      \hline
      \multirow{5}{*}{\rotatebox[origin=c]{90}{Model}}
      & UR & (3)   & \gy (1)   & (3,*) & (4) \\
      & SR & (3)   & \gy (1)   & (3,*) & (4) \\
      & CR & (2,3) & \gy (1,2) & (2,3) & (2) \\
      & DR & (2,3) & \gy (1,2) & (2,3) & (2) \\
      & TR & (3)   & \gy (1)   & (3,*) & (4) \\
      \hline
    \end{tabular}
  \end{minipage} %
  \hfill
  \begin{minipage}{0.53\textwidth}
    \footnotesize
    \begin{itemize}[noitemsep, leftmargin=*]
      \item[(1)] Not possible as CRs not reachable in real system due to readjustment in surrounding DRs.
      \item[(2)] Not possible as plans leading to CRs or DRs in the model are discarded during synthesis.
      \item[(3)] Safe due to trivial safety of SRs and DRs.
      \item[(4)] TR was safely reached and the process finishes successfully.
      \item[(*)] Could affect termination (due to infinitely repeated readjustments), but is solved by safety margins.
    \end{itemize}
  \end{minipage}
  \caption{Case distinction on regions for safety proof. (Note: Column for unknown regions (UR) in real system is left out, as each state of the real system is either known in advance or, when reached, directly classified as SR, CR, or DR based on measured data. Furthermore, URs in the model are treated as ``safe'' during strategy synthesis until they are further classified.)}
  \label{fig:region-safety}
\end{figure}

\begin{theorem}[Safety]\label{thm:safety}
 A motion plan classified as safe via offline strategy synthesis is indeed safe under the momentary system assumptions (``local safety'').  The system never reaches an unsafe state via online strategy synthesis (``global safety'').
\end{theorem}
Local safety follows directly from the soundness of the underlying strategy synthesizer. Global safety follows from the case distinction in \cref{fig:region-safety} if one can show that the safety margins are determined so, that deviations from the model cannot lead the real system into known CRs and DRs.

The needle steering application indicates that the system assumptions are reasonable:
We normally observe an increasing force when moving to another -- possibly critical -- tissue type due to deformations, which allows detection of upcoming CRs.
Spanning around $5 - 6mm$, the deformation section is larger than the measurable step size of the used sensors, which provide new force and position data with a frequency of $150Hz$ (i.e., every $33.35{\mu}m$ of needle progress at a speed of $5mm/s$) with a conservative maximum error of $3mm$;
the DRs are thus always measurable.
Note that the error is mostly static and can thus be accounted for, unless optical tracking fails, e.g., due to air bubbles or surface reflections in gelatin, which may be prevented by ultrasound measurements in the future.
Finally, rollbacks are possible as needle pullbacks will always follow the inversed insertion path, whose safety we discovered already.

Rollbacks increase the chance of finding safe plans but the OnSS algorithm does not guarantee that every safe plan is discovered. At the same time, termination of OnSS becomes non-trivial in the presence of rollbacks since one has to show that only a finite number of DRs is added. Due to space restrictions, we just state the following property without further proof.
\begin{theorem}[Incompleteness and termination]\label{thm:incompleteness}
 OnSS based on safety margins and on-the-fly discovered URs is incomplete. The OnSS instance for the needle-steering problem terminates.
\end{theorem}

\vspace{-6pt}
\paragraph{Action Plan Optimization} \label{subsec:optimization}
In general, one can distinguish between \textit{hard (H) and soft (S) requirements} for concrete action plans.
In case of needle steering, the following requirements are given:
\begin{itemize}[noitemsep, leftmargin=*]
  \begin{minipage}[t]{0.34\linewidth}
    \item \textbf{H1}: The target region is reached
    \item \textbf{S1}: As few rotations as possible are needed
    \item \textbf{S2}: The path is as short as possible
  \end{minipage}
  \hspace{1cm}
  \begin{minipage}[t]{0.58\linewidth}
    \item \textbf{H2}: No critical region (e.g., critical tissue) is pierced
    \item \textbf{S3}: The critical regions are circumvented most spaciously
    \item \textbf{S4}: The path needs the fewest amount of readjustments
    \item \textbf{S5}: The target center is reached as close as possible
  \end{minipage}
\end{itemize}

A synthesized strategy usually contains more than one possible action plan, which all automatically satisfy the binary hard requirements H1 and H2 regarding reachability and safety, respectively, based on the reachability query \texttt{EF StateChecker.Final\_TR\_Reached} and critical paths leading to deadlocks before.
The soft requirements, in contrast, are continuous by nature, and their satisfaction differs for the accepted action plans.
While some soft requirements are directly connected (e.g., S3 and S4), others usually contradict each other (e.g., S2 and S3), so that no universal optimum can be found.
Therefore, a fixed weighting, i.e., cost assignment, of the soft requirements turns the action plan choice into an optimization problem.
Depending on whether the system is discrete (using fixed time steps) or continuous (using ordinary differential equations), the costs assigned to the actions and distances in the model can be implemented either as integer variables incremented by cost deltas, or hybrid clocks, respectively.

The concrete cost values need to be provided a-priori.
For needle steering, one source of knowledge for cost weighting are the underlying biological aspects.
While the precise cost vector is still subject to ongoing research, plausible assumptions for cost assignments are that needle rotations inside the tissue impose more damage than a slightly longer path, and that -- in the scope of ``safe'' regions -- readjustments impose the highest damage.

%%%%%%%%%%%%%%%%%%%%%%%%%%%%%%%%%%%%%%%%%%%%%%%%%%%%%%%%%%%%%%%%%%%%%%%%%%%%%%%%
%%% Experiments %%%
%%%%%%%%%%%%%%%%%%%%%%%%%%%%%%%%%%%%%%%%%%%%%%%%%%%%%%%%%%%%%%%%%%%%%%%%%%%%%%%%
\section{Experiments} \label{sec:experiments}
\begin{figure}[t]
  \hspace*{\fill}%
  \subfloat[System setup]{
  \begin{minipage}[b]{0.32\textwidth}
     \centering
     \includegraphics[width=\textwidth]{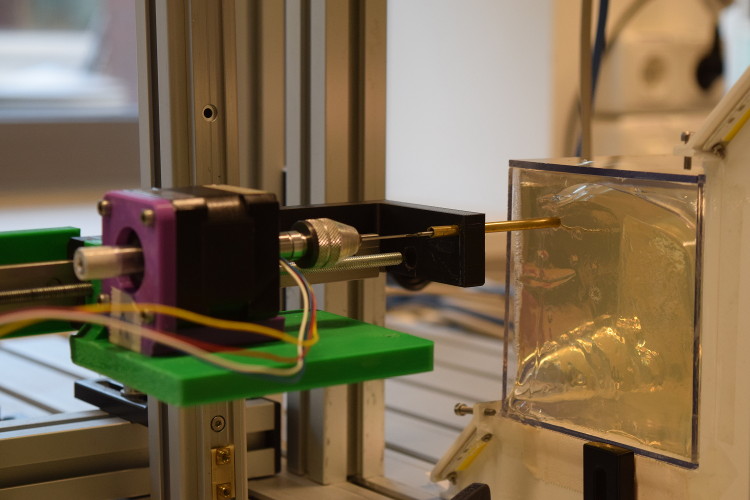}
     \label{subfig:system-setup}
  \end{minipage}}
  \hfill
  \hspace*{\fill}%
  \subfloat[Initial strategy]{
	\begin{minipage}[b]{0.32\textwidth}
	   \centering
	   \includegraphics[width=\textwidth, draft=false]{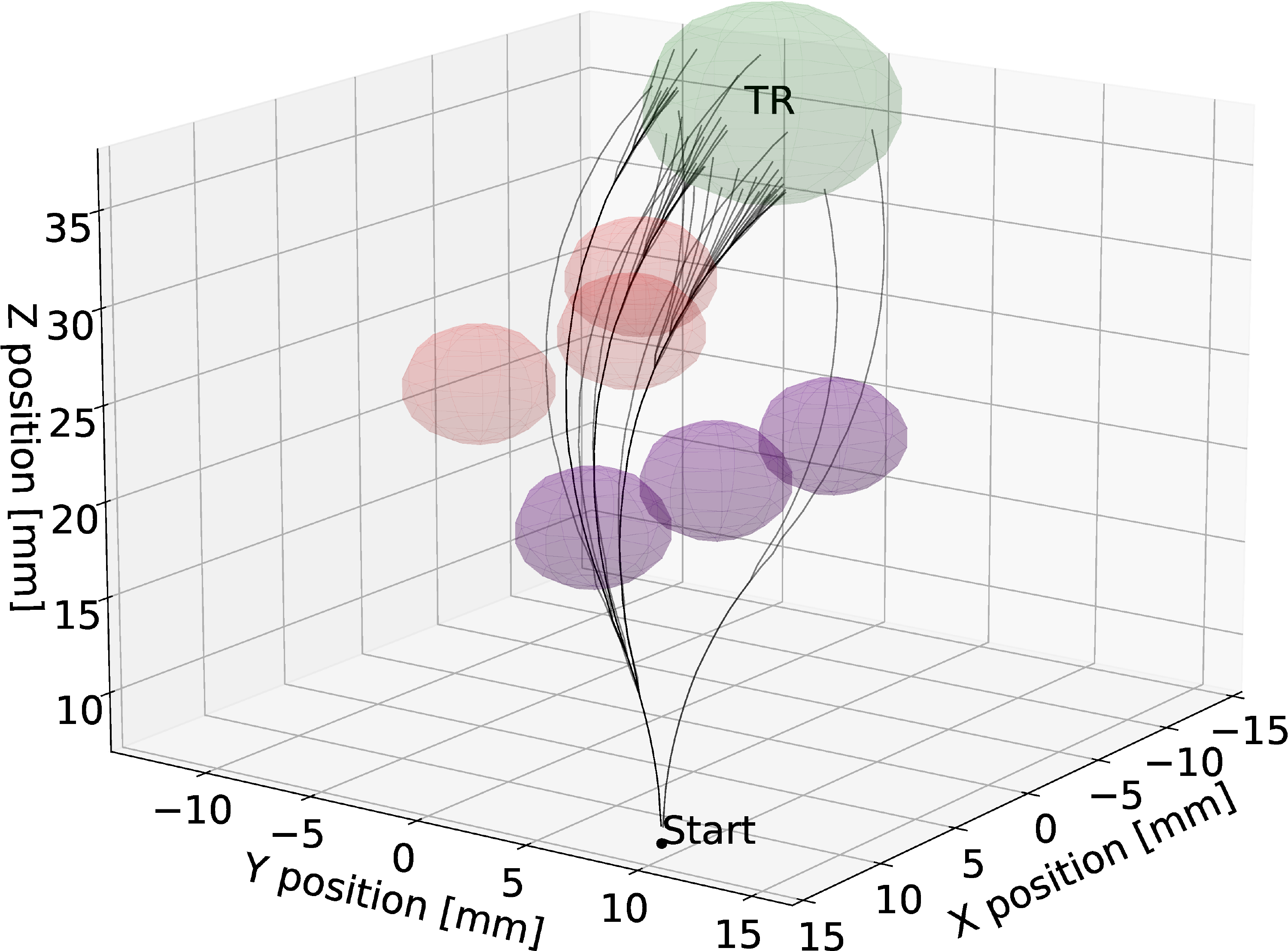}
     \label{subfig:exp-initial-motion-plans}
	\end{minipage}}
  \hfill
  \subfloat[Final needle trace]{
	\begin{minipage}[b]{0.32\textwidth}
	   \centering
	   \includegraphics[width=\textwidth, draft=false]{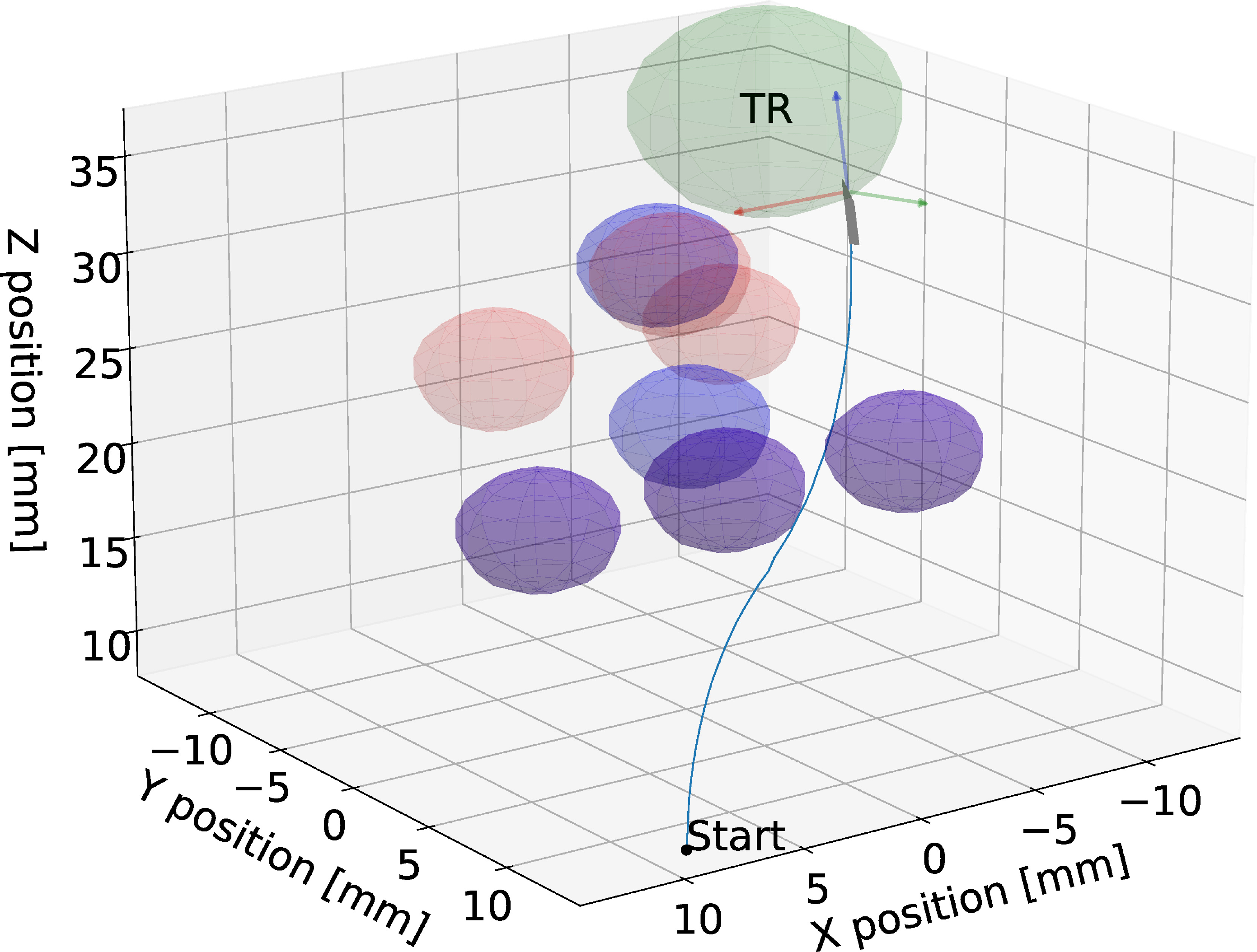}
     \label{subfig:exp-final-trace}
	\end{minipage}}
  \hspace*{\fill}%
  \caption{Experiment setup and needle motion plans with initially unknown (red), known, and discovered (both blue) CRs.}
  \label{fig:example-experiment}
\end{figure}

Following the model structure introduced in \cref{subsec:model-and-workflow}, we implemented a needle steering model in \textit{Uppaal Stratego}.
We developed an online strategy synthesis framework in \textit{Python}, which then uses verification queries on the frequently updated model for the individual synthesis steps.
We conducted two types of experiments based on randomly generated (\textbf{Experiment 1}) and real-measured reference needle paths (\textbf{Experiment 2}), where the setup shown in \cref{subfig:system-setup} is used for the latter.
The reference paths are used to randomly place target and critical regions on and around that path, respectively, as base for each invidual experiment run;
that way, we ensure that at least one safe motion plan towards the target region exists initially, so that the experiment is not aborted immediately.
An experiment run \textit{failed} if the starting state is re-reached via readjustments, or if the target region is not reached within $2$ minutes.

The experiment runs differ in the number ($\{0, 1, 2, \mathbf{5}, 10, 20\}$) and size ($\{1mm, 2mm, \mathbf{3mm}, 4mm,$ $5mm, 10mm\}$) of CRs, assumed size of identified CRs ($\{1mm, 2mm, \mathbf{3mm}, 4mm, 5mm, 10mm\}$),
distance of the TR along the reference trace ($\{10mm, 20mm, \mathbf{30mm}, 40mm, 50mm\}$), and percentage of initially known CRs ($\{\mathbf{0\%}, 20\%, 40\%, 60\%, 80\%, 100\%\}$).
One parameter is changed at a time, and the bold values are used for the other parameters that remain unchanged.
For each of the $29$ parametrizations derived hereby, we perform $20$ experiment runs, i.e., $580$ needle steering runs in total.

\begin{table}[b]
  \centering
  \begin{tabular}[t]{lccccc}
    \hline
    & TR Reach & Readjustments & Synthesis Time & Overall Time \\
    \hline
    Experiment 1 & $88.28\%$ & $(0.00,0.57,10.00)$ & $(0.04s,2.37s,5.18s)$ & $(7.42s,20.45s,85.82s)$ \\
    Experiment 2 & $68.39\%$ & $(0.00,2.94,19.00)$ & $(0.04s,2.34s,6.99s)$ & $(7.61s,36.03s,119.50s)$ \\
    \hline
  \end{tabular}
  \caption{Experiment results for each measure with (min,avg,max) data}
  \label{fig:general-result-table}
\end{table}

\cref{subfig:exp-initial-motion-plans} and \cref{subfig:exp-final-trace} exemplarily show the set of initially (offline) synthesized motion plans as well as the actually (online) traversed path for a single experiment run.
Furthermore, the measurement results are shown in \cref{fig:general-result-table}, and highlight four aspects:
First, for experiments 1 and 2, the success rates of reaching a target are between $68 - 88\%$;
the rates stay below $100\%$ because of incompleteness (cf. \cref{subsec:safety}) and imprecision of the underlying geometric model (only experiment 2).
Second, the number of readjustments on average is comparatively low ($0.57$ and $2.94$, respectively); some runs even need no readjustment at all.
Third, the synthesis times lie between $0.04s$ and $6.99s$, which appear acceptable for an online application.
Fourth, the successful runs took between $20.45s$ and $36.03s$ on average, so that the complete procedure (from insertion to target reaching) can be finished in reasonable time.

%%%%%%%%%%%%%%%%%%%%%%%%%%%%%%%%%%%%%%%%%%%%%%%%%%%%%%%%%%%%%%%%%%%%%%%%%%%%%%%%
%%% Conclusion %%%
%%%%%%%%%%%%%%%%%%%%%%%%%%%%%%%%%%%%%%%%%%%%%%%%%%%%%%%%%%%%%%%%%%%%%%%%%%%%%%%%
\section{Conclusion and Future Work} \label{sec:conclusion}
We presented an online workflow to realize controller synthesis for real-timed games in (changing) environments of partial knowledge.
We classify the environment into unknown, safe, critical, detection, and target regions and ensure safe action planning to reach a target state by periodically validating and updating our model.
We applied our approach to medical needle steering.

In future work, we plan to replace the geometric approximation of needle motion by a physically accurate model and increase the ``hostility'' of the environment by permitting displacement of critical regions due to respiration, deformed tissue layers (which affect the needle tip speed), and inhomogeneous properties of multi-layered tissue.
Work on human-robot collaboration \cite{Askarpour2020} may then provide a starting point for formal modeling and integration of more advanced human behavior.

%%%%%%%%%%%%%%%%%%%%%%%%%%%%%%%%%%%%%%%%%%%%%%%%%%%%%%%%%%%%%%%%%%%%%%%%%%%%%%%%
%%% BIBLIOGRAPHY %%%
%%%%%%%%%%%%%%%%%%%%%%%%%%%%%%%%%%%%%%%%%%%%%%%%%%%%%%%%%%%%%%%%%%%%%%%%%%%%%%%%
\bibliographystyle{eptcs}
\bibliography{bib/references}

\begin{thebibliography}{10}
\providecommand{\bibitemdeclare}[2]{}
\providecommand{\surnamestart}{}
\providecommand{\surnameend}{}
\providecommand{\urlprefix}{Available at }
\providecommand{\url}[1]{\texttt{#1}}
\providecommand{\href}[2]{\texttt{#2}}
\providecommand{\urlalt}[2]{\href{#1}{#2}}
\providecommand{\doi}[1]{doi:\urlalt{http://dx.doi.org/#1}{#1}}
\providecommand{\eprint}[1]{arXiv:\urlalt{https://arxiv.org/abs/#1}{#1}}
\providecommand{\bibinfo}[2]{#2}

\bibitemdeclare{inproceedings}{Altisen2002}
\bibitem{Altisen2002}
\bibinfo{author}{Karine \surnamestart Altisen\surnameend} \&
  \bibinfo{author}{Stavros \surnamestart Tripakis\surnameend}
  (\bibinfo{year}{2002}): \emph{\bibinfo{title}{Tools for Controller Synthesis
  of Timed Systems}}.
\newblock In: {\sl \bibinfo{booktitle}{Proceedings 2nd Workshop on Real-Time
  Tools}}.

\bibitemdeclare{article}{Alur1994}
\bibitem{Alur1994}
\bibinfo{author}{Rajeev \surnamestart Alur\surnameend} \&
  \bibinfo{author}{David~L. \surnamestart Dill\surnameend}
  (\bibinfo{year}{1994}): \emph{\bibinfo{title}{{A Theory of Timed Automata}}}.
\newblock {\sl \bibinfo{journal}{Theoretical Computer Science}}
  \bibinfo{volume}{126}, pp. \bibinfo{pages}{183--235},
  \doi{10.1016/0304-3975(94)90010-8}.

\bibitemdeclare{inproceedings}{Asarin1995}
\bibitem{Asarin1995}
\bibinfo{author}{Eugene \surnamestart Asarin\surnameend}, \bibinfo{author}{Oded
  \surnamestart Maler\surnameend} \& \bibinfo{author}{Amir \surnamestart
  Pnueli\surnameend} (\bibinfo{year}{1995}): \emph{\bibinfo{title}{{Symbolic
  controller synthesis for discrete and timed systems}}}.
\newblock In: {\sl \bibinfo{booktitle}{Hybrid Systems II}}, pp.
  \bibinfo{pages}{1--20}, \doi{10.1007/3-540-60472-3\_1}.

\bibitemdeclare{inproceedings}{Askarpour2020}
\bibitem{Askarpour2020}
\bibinfo{author}{Mehrnoosh \surnamestart Askarpour\surnameend}
  (\bibinfo{year}{2020}): \emph{\bibinfo{title}{How to Formally Model Human in
  Collaborative Robotics}}.
\newblock In: {\sl \bibinfo{booktitle}{{\rm Proceedings Second Workshop on}
  Formal Methods for Autonomous Systems}}, \bibinfo{volume}{329}, pp.
  \bibinfo{pages}{1--14}, \doi{10.4204/EPTCS.329.1}.

\bibitemdeclare{article}{Bacci2021}
\bibitem{Bacci2021}
\bibinfo{author}{Giovanni \surnamestart Bacci\surnameend},
  \bibinfo{author}{Patricia \surnamestart Bouyer\surnameend},
  \bibinfo{author}{Uli \surnamestart Fahrenberg\surnameend},
  \bibinfo{author}{Kim~G. \surnamestart Larsen\surnameend},
  \bibinfo{author}{Nicolas \surnamestart Markey\surnameend} \&
  \bibinfo{author}{Pierre-Alain \surnamestart Reynier\surnameend}
  (\bibinfo{year}{2021}): \emph{\bibinfo{title}{Optimal and robust controller
  synthesis using energy timed automata with uncertainty}}.
\newblock {\sl \bibinfo{journal}{Formal Aspects of Computing}}
  \bibinfo{volume}{33}(\bibinfo{number}{1}), pp. \bibinfo{pages}{3--25},
  \doi{10.1007/s00165-020-00521-4}.

\bibitemdeclare{inproceedings}{Bertrand2012}
\bibitem{Bertrand2012}
\bibinfo{author}{Nathalie \surnamestart Bertrand\surnameend} \&
  \bibinfo{author}{Sven \surnamestart Schewe\surnameend}
  (\bibinfo{year}{2012}): \emph{\bibinfo{title}{{Playing Optimally on Timed
  Automata with Random Delays}}}.
\newblock In: {\sl \bibinfo{booktitle}{{FORMATS 2012} - Formal Modeling and
  Analysis of Timed Systems}}, pp. \bibinfo{pages}{43--58},
  \doi{10.1007/978-3-642-33365-1\_5}.

\bibitemdeclare{article}{Bilgram2021}
\bibitem{Bilgram2021}
\bibinfo{author}{Alexander \surnamestart Bilgram\surnameend},
  \bibinfo{author}{Emil \surnamestart Ernstsen\surnameend},
  \bibinfo{author}{Peter \surnamestart Greve\surnameend},
  \bibinfo{author}{Harry \surnamestart Lahrmann\surnameend},
  \bibinfo{author}{{Kim G.} \surnamestart Larsen\surnameend},
  \bibinfo{author}{Marco \surnamestart Mu{\~n}iz\surnameend},
  \bibinfo{author}{Peter \surnamestart Taankvist\surnameend} \&
  \bibinfo{author}{Thomas \surnamestart Pedersen\surnameend}
  (\bibinfo{year}{2021}): \emph{\bibinfo{title}{Online and Proactive Vehicle
  Rerouting with Uppaal Stratego}}.
\newblock {\sl \bibinfo{journal}{Transportation Research Record}},
  \doi{10.1177/03611981211000348}.

\bibitemdeclare{inproceedings}{Bischopink2020}
\bibitem{Bischopink2020}
\bibinfo{author}{Christopher \surnamestart Bischopink\surnameend} \&
  \bibinfo{author}{Maike \surnamestart Schwammberger\surnameend}
  (\bibinfo{year}{2020}): \emph{\bibinfo{title}{Verification of Fair
  Controllers for Urban Traffic Manoeuvres at Intersections}}.
\newblock In: {\sl \bibinfo{booktitle}{FM 2019 - Formal Methods. FM 2019
  International Workshops}}, pp. \bibinfo{pages}{249--264},
  \doi{10.1007/978-3-030-54994-7\_18}.

\bibitemdeclare{inproceedings}{Bouyer2004}
\bibitem{Bouyer2004}
\bibinfo{author}{Patricia \surnamestart Bouyer\surnameend},
  \bibinfo{author}{Franck \surnamestart Cassez\surnameend},
  \bibinfo{author}{Emmanuel \surnamestart Fleury\surnameend} \&
  \bibinfo{author}{Kim~Guldstrand \surnamestart Larsen\surnameend}
  (\bibinfo{year}{2004}): \emph{\bibinfo{title}{{Optimal Strategies in Priced
  Timed Game Automata}}}.
\newblock In: {\sl \bibinfo{booktitle}{FSTTCS 2004 - Foundations of Software
  Technology and Theoretical Computer Science}}, pp. \bibinfo{pages}{148--160},
  \doi{10.1007/978-3-540-30538-5\_13}.

\bibitemdeclare{inproceedings}{Bouyer2009}
\bibitem{Bouyer2009}
\bibinfo{author}{Patricia \surnamestart Bouyer\surnameend} \&
  \bibinfo{author}{Vojt{\v{e}}ch \surnamestart Forejt\surnameend}
  (\bibinfo{year}{2009}): \emph{\bibinfo{title}{{Reachability in Stochastic
  Timed Games}}}.
\newblock In: {\sl \bibinfo{booktitle}{ICALP 2009 - Automata, Languages, and
  Programming}}, pp. \bibinfo{pages}{103--114},
  \doi{10.1007/978-3-642-02930-1\_9}.

\bibitemdeclare{inproceedings}{Cassez2005}
\bibitem{Cassez2005}
\bibinfo{author}{Franck \surnamestart Cassez\surnameend},
  \bibinfo{author}{Alexandre \surnamestart David\surnameend},
  \bibinfo{author}{Emmanuel \surnamestart Fleury\surnameend},
  \bibinfo{author}{Kim~Guldstrand \surnamestart Larsen\surnameend} \&
  \bibinfo{author}{Didier \surnamestart Lime\surnameend}
  (\bibinfo{year}{2005}): \emph{\bibinfo{title}{{Efficient On-the-Fly
  Algorithms for the Analysis of Timed Games}}}.
\newblock In: {\sl \bibinfo{booktitle}{{CONCUR} 2005 - Concurrency Theory}},
  pp. \bibinfo{pages}{66--80}, \doi{10.1007/11539452\_9}.

\bibitemdeclare{inproceedings}{Cassez2007}
\bibitem{Cassez2007}
\bibinfo{author}{Franck \surnamestart Cassez\surnameend},
  \bibinfo{author}{Alexandre \surnamestart David\surnameend},
  \bibinfo{author}{Kim~G. \surnamestart Larsen\surnameend},
  \bibinfo{author}{Didier \surnamestart Lime\surnameend} \&
  \bibinfo{author}{Jean-Fran{\c{c}}ois \surnamestart Raskin\surnameend}
  (\bibinfo{year}{2007}): \emph{\bibinfo{title}{Timed Control with Observation
  Based and Stuttering Invariant Strategies}}.
\newblock In: {\sl \bibinfo{booktitle}{ATVA 2007 - Automated Technology for
  Verification and Analysis}}, pp. \bibinfo{pages}{192--206},
  \doi{10.1007/978-3-540-75596-8\_15}.

\bibitemdeclare{inproceedings}{David2015}
\bibitem{David2015}
\bibinfo{author}{Alexandre \surnamestart David\surnameend},
  \bibinfo{author}{Peter~Gj{\o}l \surnamestart Jensen\surnameend},
  \bibinfo{author}{Kim~Guldstrand \surnamestart Larsen\surnameend},
  \bibinfo{author}{Marius \surnamestart Miku{\v{c}}ionis\surnameend} \&
  \bibinfo{author}{Jakob~Haahr \surnamestart Taankvist\surnameend}
  (\bibinfo{year}{2015}): \emph{\bibinfo{title}{Uppaal Stratego}}.
\newblock In: {\sl \bibinfo{booktitle}{TACAS 2015 - Tools and Algorithms for
  the Construction and Analysis of Systems}}, pp. \bibinfo{pages}{206--211},
  \doi{10.1007/978-3-662-46681-0\_16}.

\bibitemdeclare{inproceedings}{David2009}
\bibitem{David2009}
\bibinfo{author}{Alexandre \surnamestart David\surnameend},
  \bibinfo{author}{{Kim Guldstrand} \surnamestart Larsen\surnameend} \&
  \bibinfo{author}{Thomas \surnamestart Chatain\surnameend}
  (\bibinfo{year}{2009}): \emph{\bibinfo{title}{Playing Games with Timed
  Games}}.
\newblock In: {\sl \bibinfo{booktitle}{ADHS 2003 - IFAC Proceedings Volumes}},
  pp. \bibinfo{pages}{238--243}, \doi{10.3182/20090916-3-ES-3003.00042}.

\bibitemdeclare{inproceedings}{Daws1996}
\bibitem{Daws1996}
\bibinfo{author}{C.~\surnamestart Daws\surnameend},
  \bibinfo{author}{A.~\surnamestart Olivero\surnameend},
  \bibinfo{author}{S.~\surnamestart Tripakis\surnameend} \&
  \bibinfo{author}{S.~\surnamestart Yovine\surnameend} (\bibinfo{year}{1996}):
  \emph{\bibinfo{title}{{The tool Kronos}}}.
\newblock In: {\sl \bibinfo{booktitle}{Hybrid Systems III}}, pp.
  \bibinfo{pages}{208--219}, \doi{10.1007/BFb0020947}.

\bibitemdeclare{article}{Ferrando2021}
\bibitem{Ferrando2021}
\bibinfo{author}{Angelo \surnamestart Ferrando\surnameend},
  \bibinfo{author}{Louise~A. \surnamestart Dennis\surnameend},
  \bibinfo{author}{Rafael~C. \surnamestart Cardoso\surnameend},
  \bibinfo{author}{Michael \surnamestart Fisher\surnameend},
  \bibinfo{author}{Davide \surnamestart Ancona\surnameend} \&
  \bibinfo{author}{Viviana \surnamestart Mascardi\surnameend}
  (\bibinfo{year}{2021}): \emph{\bibinfo{title}{Toward a Holistic Approach to
  Verification and Validation of Autonomous Cognitive Systems}}.
\newblock {\sl \bibinfo{journal}{ACM Transactions on Software Engineering and
  Methodology}} \bibinfo{volume}{30}(\bibinfo{number}{4}),
  \doi{10.1145/3447246}.

\bibitemdeclare{inproceedings}{Finkbeiner2012}
\bibitem{Finkbeiner2012}
\bibinfo{author}{Bernd \surnamestart Finkbeiner\surnameend} \&
  \bibinfo{author}{Hans-J{\"o}rg \surnamestart Peter\surnameend}
  (\bibinfo{year}{2012}): \emph{\bibinfo{title}{Template-Based Controller
  Synthesis for Timed Systems}}.
\newblock In: {\sl \bibinfo{booktitle}{TACAS 2012 - Tools and Algorithms for
  the Construction and Analysis of Systems}}, pp. \bibinfo{pages}{392--406},
  \doi{10.1007/978-3-642-28756-5\_27}.

\bibitemdeclare{article}{Gleirscher2021}
\bibitem{Gleirscher2021}
\bibinfo{author}{Mario \surnamestart Gleirscher\surnameend},
  \bibinfo{author}{Radu \surnamestart Calinescu\surnameend} \&
  \bibinfo{author}{Jim \surnamestart Woodcock\surnameend}
  (\bibinfo{year}{2021}): \emph{\bibinfo{title}{RiskStructures: A Design
  Algebra for Risk-Aware Machines}}.
\newblock {\sl \bibinfo{journal}{Formal Aspects of Computing}}
  \bibinfo{volume}{33}, pp. \bibinfo{pages}{763--802},
  \doi{10.1007/s00165-021-00545-4}.

\bibitemdeclare{inproceedings}{Larsen2018}
\bibitem{Larsen2018}
\bibinfo{author}{Kim~G. \surnamestart Larsen\surnameend},
  \bibinfo{author}{Florian \surnamestart Lorber\surnameend} \&
  \bibinfo{author}{Brian \surnamestart Nielsen\surnameend}
  (\bibinfo{year}{2018}): \emph{\bibinfo{title}{20 Years of UPPAAL Enabled
  Industrial Model-Based Validation and Beyond}}.
\newblock In: {\sl \bibinfo{booktitle}{Leveraging Applications of Formal
  Methods, Verification and Validation. Industrial Practice}}, pp.
  \bibinfo{pages}{212--229}, \doi{10.1007/978-3-030-03427-6\_18}.

\bibitemdeclare{inproceedings}{Larsen2016}
\bibitem{Larsen2016}
\bibinfo{author}{Kim~G. \surnamestart Larsen\surnameend},
  \bibinfo{author}{Marius \surnamestart Miku{\v{c}}ionis\surnameend},
  \bibinfo{author}{Marco \surnamestart Mu{\~{n}}iz\surnameend},
  \bibinfo{author}{Ji{\v{r}}{\'i} \surnamestart Srba\surnameend} \&
  \bibinfo{author}{Jakob~Haahr \surnamestart Taankvist\surnameend}
  (\bibinfo{year}{2016}): \emph{\bibinfo{title}{{Online and Compositional
  Learning of Controllers with Application to Floor Heating}}}.
\newblock In: {\sl \bibinfo{booktitle}{Tools and Algorithms for the
  Construction and Analysis of Systems}}, pp. \bibinfo{pages}{244--259},
  \doi{10.1007/978-3-662-49674-9\_14}.

\bibitemdeclare{article}{LuiY2017}
\bibitem{LuiY2017}
\bibinfo{author}{Yang \surnamestart Liu\surnameend},
  \bibinfo{author}{Yu~\surnamestart Peng\surnameend}, \bibinfo{author}{Bailing
  \surnamestart Wang\surnameend}, \bibinfo{author}{Sirui \surnamestart
  Yao\surnameend} \& \bibinfo{author}{Zihe \surnamestart Liu\surnameend}
  (\bibinfo{year}{2017}): \emph{\bibinfo{title}{Review on cyber-physical
  systems}}.
\newblock {\sl \bibinfo{journal}{IEEE/CAA Journal of Automatica Sinica}}
  \bibinfo{volume}{4}(\bibinfo{number}{1}), pp. \bibinfo{pages}{27--40},
  \doi{10.1109/JAS.2017.7510349}.

\bibitemdeclare{inproceedings}{Maler1995}
\bibitem{Maler1995}
\bibinfo{author}{Oded \surnamestart Maler\surnameend}, \bibinfo{author}{Amir
  \surnamestart Pnueli\surnameend} \& \bibinfo{author}{Joseph \surnamestart
  Sifakis\surnameend} (\bibinfo{year}{1995}): \emph{\bibinfo{title}{{On the
  synthesis of discrete controllers for timed systems}}}.
\newblock In: {\sl \bibinfo{booktitle}{STACS 95 - Annual Symposium on
  Theoretical Aspects of Computer Science}}, pp. \bibinfo{pages}{229--242},
  \doi{10.1007/3-540-59042-0\_76}.

\bibitemdeclare{inproceedings}{Peter2011}
\bibitem{Peter2011}
\bibinfo{author}{Hans-J{\"o}rg \surnamestart Peter\surnameend},
  \bibinfo{author}{R{\"u}diger \surnamestart Ehlers\surnameend} \&
  \bibinfo{author}{Robert \surnamestart Mattm{\"u}ller\surnameend}
  (\bibinfo{year}{2011}): \emph{\bibinfo{title}{{Synthia: Verification and
  Synthesis for Timed Automata}}}.
\newblock In: {\sl \bibinfo{booktitle}{CAV 2011 - Computer Aided
  Verification}}, pp. \bibinfo{pages}{649--655},
  \doi{10.1007/978-3-642-22110-1\_52}.

\bibitemdeclare{article}{Rogalla2020}
\bibitem{Rogalla2020}
\bibinfo{author}{Antje \surnamestart Rogalla\surnameend},
  \bibinfo{author}{Sascha \surnamestart Lehmann\surnameend},
  \bibinfo{author}{Maximilian \surnamestart Neidhardt\surnameend},
  \bibinfo{author}{Johanna \surnamestart Sprenger\surnameend},
  \bibinfo{author}{Marcel \surnamestart Bengs\surnameend},
  \bibinfo{author}{Alexander \surnamestart Schlaefer\surnameend} \&
  \bibinfo{author}{Sibylle \surnamestart Schupp\surnameend}
  (\bibinfo{year}{2020}): \emph{\bibinfo{title}{{Synthesizing Strategies for
  Needle Steering in Gelatin Phantoms}}}.
\newblock {\sl \bibinfo{journal}{Electronic Proceedings in Theoretical Computer
  Science}} \bibinfo{volume}{316}, pp. \bibinfo{pages}{261--274},
  \doi{10.4204/eptcs.316.10}.

\bibitemdeclare{article}{Webster2006}
\bibitem{Webster2006}
\bibinfo{author}{Robert~J. \surnamestart Webster\surnameend},
  \bibinfo{author}{Jin~Seob \surnamestart Kim\surnameend},
  \bibinfo{author}{Noah~J. \surnamestart Cowan\surnameend},
  \bibinfo{author}{Gregory~S. \surnamestart Chirikjian\surnameend} \&
  \bibinfo{author}{Allison~M. \surnamestart Okamura\surnameend}
  (\bibinfo{year}{2006}): \emph{\bibinfo{title}{{Nonholonomic Modeling of
  Needle Steering}}}.
\newblock {\sl \bibinfo{journal}{International Journal of Robotics Research}}
  \bibinfo{volume}{25}(\bibinfo{number}{5-6}), pp. \bibinfo{pages}{509--525},
  \doi{10.1177/0278364906065388}.

\end{thebibliography}
% \nocite{*}

\end{document}